\documentclass[11pt]{iopart}

\usepackage{amssymb}
\usepackage{graphicx}
\usepackage{multirow}
\usepackage{booktabs,siunitx}

\begin{document}

\title[]{Existence of inter coupled structural, electronic and magnetic states in Sm$ _{2} $NiMnO$ _{6} $ double perovskite}

\author{
S. Majumder$^{a}$,
\
M. Tripathi$^{a}$,
\
D. O. de Souza$^{b}$,
\
A. Sagdeo$^{c,d}$,
\
L. Olivi$^{b}$,
M. N. Singh$^{c}$,
\
S. Pal$^{c}$,
\
R. J. Choudhary$^{a,*}$
and
D. M. Phase$^{a}$
\\
$^{a}$UGC DAE Consortium for Scientific Research, Indore 452001, India\\
$^{b}$Elettra Sicrotrone Trieste S.C.p.A., SS 14-km 163.5, 34149 Basovizza, Italy\\
$^{c}$HXAL, SUS, Raja Ramanna Centre for Advanced Technology, Indore 452013, India\\
$^{d}$Homi Bhabha National Institute, Anushakti nagar, Mumbai 400 094, India\\}
\ead{$^{*}$ram@csr.res.in}

\begin{abstract}
Coupling between different interactions allows to control physical aspects in multifunctional materials by perturbing any of the degrees of freedom. Here, we aim to probe the correlation among structural, electronic and magnetic observables of Sm$ _{2} $NiMnO$ _{6} $ ferromagnetic insulator double perovskite. Our employed methodology includes thermal evolution of synchrotron X-ray diffraction, near edge and extended edge hard X-ray absorption spectroscopy and bulk magnetometry. The magnetic ordering in SNMO adopts two transitions, at T$ _{C} $=159.6K due to ferromagnetic arrangement of Ni-Mn sublattice and at T$ _{d} $=34.1K because of anti-parallel alignment of polarized Sm paramagnetic moments with respect to Ni-Mn network. The global as well as local crystal structure of SNMO undergoes isostructural transitions across T$ _{C} $ and T$ _{d} $, observed by means of temperature dependent variation in Ni/Mn-O, Ni-Mn bonding characters and super exchange angle in Ni-O-Mn linkage. Hybridization between Ni, Mn 3\textit{d}, O 2\textit{p} electronic states is also modified in the vicinity of magnetic transition. On the other hand, the signature of Ni/Mn anti-site disorders are evidenced from local structure and magnetization analysis. The change in crystal environments governs the magnetic response by imposing alteration in metal - ligand orbital overlap. Utilizing these complimentary probes we have found that structural, electronic and magnetic states are inter-coupled in SNMO which makes it a potential platform for technological usage. 
\end{abstract}

\section{INTRODUCTION}
In materials science there has been always a quest for the strong correlation between charge, spin and lattice degrees of freedom to have additional control over functional aspects of the system \cite{MImada1998}. Double perovskites with Ni and Mn as B-site cations A$ _{2} $NiMnO$ _{6} $ (ANMO), have huge potential in new generation quantum electronics owing to rare tunable ferromagnetic insulating ground state \cite{DChoudhury2012, NSRogadol2005, SMajumder2022prbb, SMajumder2022prbt, SMajumder2022jpcm}. In a previous theoretical study it is predicted that LNMO (A=La) holds magnetic order dependent electronic properties \cite{HDas2009}. Later, the evidences of magneto-dielectric, magneto-phononic and magneto-elastic correlations are found in LNMO \cite{DChoudhury2012, PKumar2014, DYang2019}. On the other hand, based on first principal calculations it is argued that ANMO system may have structural strain driven multiferroic characters \cite{HJZhao2014} which are also experimentally verified in strained LNMO phase \cite{RTakahashi2015}. Aforementioned reports indicate that there is an interplay in between charge, spin and lattice, which governs the intriguing properties in ANMO family. However, understanding of these possible inter-coupled multifarious interactions needs comprehensive experimental insights, which is still elusive.

Ground state calculations on ANMO family have predicted that for different A-site ions, the magnetic ordering can transform from ferromagnetic (for A=La) to E type antiferromagnetic (for A=Y and In) \cite{SKumar2010, WYi2013}. The experimental magnetic structure analysis reveals that ANMO for A=La, Nd, Sm and Y ions, it is collinear ferromagnetic \cite{NSRogadol2005, SanchezBenitez2011, HNhalil2015}; for A=Tb, Ho, Er and Tm ions, it is canted ferrimagnetic \cite{MRetuerto2015}; for A=In, it is incommensurate antiferromagnetic \cite{NTerada2015} and for A=Sc, it is collinear antiferromagnetic \cite{WYi2015}. These studies point out that magnetic ordering in ANMO is highly sensitive to chemical pressure on the lattice structure exerted by different A-site ion sizes. At room temperature, depending upon the cation arrangement there are two possible crystal symmetry for ANMO system, monoclinic \textit{P2$_{1}$/n} and orthorhombic \textit{Pbnm} \cite{SMajumder2022prbb, SanchezBenitez2011, WZYang2012}. In monoclinic structure Ni/Mn ions arrange in ideal rock-salt fashion at alternate octahedral center sites. Whereas, a random occupation between Ni/Mn ions are formed for orthorhombic case. However in practical crystal system, completely ordered or disordered situation is difficult to achieve due to comparable ionic sizes of Ni and Mn species \cite{SMajumder2022prbb, SMajumder2022prbt, SMajumder2022jpcm}. Even in highly ordered system, some fraction of disorder is present owing to anti-phase boundary formation \cite{HZGuok2008}. Different ordered and disordered coordinations have drastic role in stabilizing the magnetic phase of the system \cite{SMajumder2022prbb, SMajumder2022prbt, SMajumder2022jpcm}. The temperature effect on local and global crystal environments of ANMO is yet to be explored. On the other hand, in previous studies different interpretations about the charge states of Ni and Mn ions are found. For instance, according to Goodenough \textit{et al.} \cite{JBGoodenoughl1961} the magnetic exchange in LNMO (for A=La) is in between Ni$^{3+}$ and Mn$ ^{3+} $ ions whereas, Blasse \textit{et al.} \cite{GBlassel1965} claims for Ni$^{2+}$ and Mn$ ^{4+} $ magnetic interaction. Further more, for perovskite nikelets and manganites, temperature driven valence state transition is reported in literature \cite{MNaka2016, NMannella2008}. So, it is necessary to check the possibility of such temperature induced valence transition in ANMO system. Moreover, the thermal evolution of crystal and electronic structures holds the key to interpret the unique ferromagnetic-insulating ground state in ANMO family of materials.

Despite several research endeavors, whether the structural, electronic and magnetic observables are inter-coupled in multifuctional ANMO double perovskite or not, is still an open question. In this backdrop we aim to investigate the thermal evolution of local and global structure, electronic state and magnetic ordering in Sm$ _{2} $NiMnO$ _{6} $ (SNMO) system. Double perovskite with Sm at A-site is chosen as a representative platform of ANMO family owing to unique mixed multiplet state of rare earth Sm$ ^{3+} $ ion which can be perturbed by crystal field, exchange field and/or thermal energy \cite{KHJBuschow1974}. Here, we have utilized a comprehensive analysis of temperature dependent synchrotron radiation X-ray diffraction, hard X-ray absorption spectroscopy in near edge and extended edge regions and bulk magnetometric measurements. The findings of present work establish that in SNMO structural, electronic and magnetic observables are correlated with each other and hence physical properties can be tuned by manipulating any of these degrees of freedom. 

\section{EXPERIMENTAL DETAILS}
SNMO polycrystalline bulk sample was synthesized by solid state reaction process following the recipe as described in Ref.\cite{SMajumder2022prbb}. The temperature dependence on global crystal structure was investigated by recording powder X-ray diffraction (PXRD) scans in temperature range of 5K$ \leqslant $T$ \leqslant $300K, utilizing synchrotron radiation source at angle dispersive X-ray diffraction (ADXRD) beamline (BL-12, Indus-II, RRCAT, Indore, India). The PXRD profiles were collected using a mar345 image plate detector and obtained two-dimensional patterns were integrated to $ 2\theta $ scales using the FIT2D program \cite{APHammersley1996}. Incident X-ray wavelength was calibrated by measuring data for reference LaB$ _{6} $ sample and the estimated value was found to be $ \lambda $=0.71949$ \AA $. Structural refinements of PXRD data were carried out employing FULLPROF software package \cite{JRodriguezCarvajal1993} wherein for background simulations WINPLOTR was used and for visualization of obtained structure VESTA was used. Elemental valence state were probed by Ni and Mn \textit{K} X-ray absorption near edge spectroscopy (XANES) performed within 13K$ \leqslant $T$ \leqslant $300K. The temperature dependency of local structure was examined by Ni \textit{K} edge extended X-ray absorption fine structure (EXAFS) spectroscopy measured in temperature range of 13K$ \leqslant $T$ \leqslant $300K. The occurrence of Sm \textit{L}$ _{3} $ edge (6716eV) close to Mn \textit{K} edge (6539eV) limits the EXAFS study for Mn case. XANES and EXAFS experiments were carried out in transmission mode using ionization chamber detectors (Oxford Instruments) and hard X-ray radiation source at XAFS beamline (11.1R, Elettra-Sincrotrone, Trieste, Italy). Incident beam photon energy calibration was done by recording data for Ni and Mn reference metal foils. The intrinsic energy resolution energy resolution $ \Delta E / E $ was $ \sim $1$ \times $10$ ^{-4} $ across the explored energy regime. To obtain normalized absorption, standard normalization process were applied on XANES spectra using ATHENA program \cite{BRavel2005}. EXAFS spectra were colloected several times to check data reproducibility and merged to have better statistics. After normalization of data, standard background subtraction process were followed to extract $k$-space ($ \chi(k) $) EXAFS signal utilizing AUTOBK algorithm \cite{MNewville1993} in ATHENA package. In order to obtain EXAFS oscillation in $ R $-space ($ \chi(R) $), Fourier transformations of the $k$-space signal (within selected $k$ range) were computed. Local structural analysis of EXAFS data were carried out using ARTEMIS software package wherein to generate theoretical standers, ATOMS and FEFF6 programs \cite{BRavel2001, JJRehr200009} were implemented. For each coordination shell (assume $ i $ number of shells) the average coordination distance ($ R_{i} $) and corresponding mean-square relative displacement factor ($ \sigma_{i}^{2} $) were computed from standard EXAFS expression \cite{JJRehr200009}, wherein all partial contributions from individual scattering paths ($ i $) were summed over to simulate model EXAFS spectra. The dc magnetic response were investigated employing MPMS 7-Tesla SQUID-VSM (Quantum Design Inc., USA) system. Prior to measurement the sample was warmed well above corresponding magnetic transition temperatures to erase previous (if any) magnetic history and standard \textit{de}-\textit{Gaussing} protocol was applied to nullify the trapped (if any) magnetic field in the magnetometer superconducting coil. Estimated average magnetic moment sensitivity was $ \sim $10$ ^{-8} $emu. 

\begin{figure}[t]
\centering
\includegraphics[angle=0,width=1.0\textwidth]{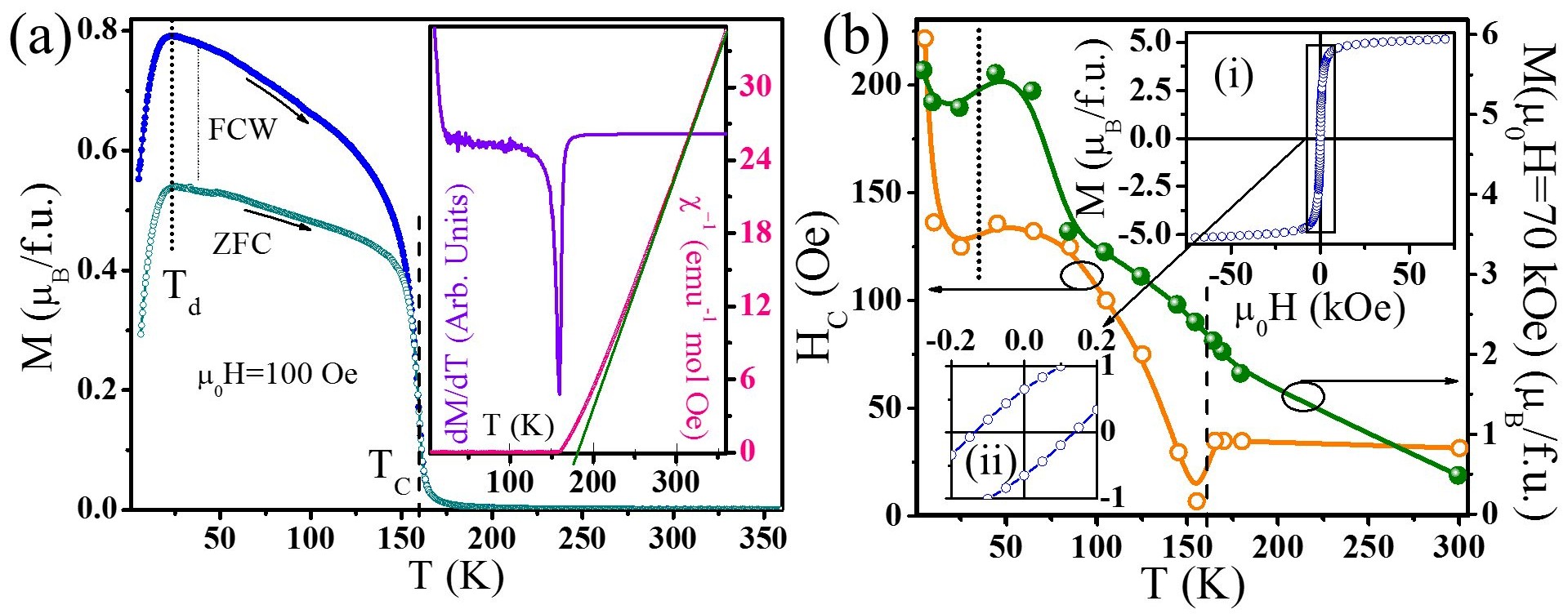}
\caption{(a): Magnetization as a function of temperature M(T), measured in zero filed cooled warming and field cooled warming cycles with applied magnetic field of $ \mu_{0}H $=100Oe. Inset shows (left panel): first order temperature derivative of magnetization dM(T)/dT (violet curve) and (right panel): inverse susceptibility 1/$ \chi $(T) observed data (pink curve) along with Curie–Weiss fitting (olive line). Thermal evolution of (b left panel): coercivity and (b right panel): magnetic moment recorded with $ \mu_{0} $H=70kOe. Inset (i): Isothermal magnetization as a function of magnetic field M(H), acquired at T=10K. Inset (ii): Enlarged view across low field region of T=10K M(H) curve. The vertical dashed and dotted lines correspond to magnetic transition temperatures T$ _{C} $ and T$ _{d} $, respectively.}\label{mtmh}
\end{figure}

\section{RESULTS AND DISCUSSION} 
The temperature dependent magnetization M(T), measured in typical zero field cooled (ZFC) warming and field cooled warming (FCW) protocols, in presence of applied dc magnetic field $ \mu_{0} $H=100Oe, are presented in Fig. \ref{mtmh}(a). Upon cooling from room temperature, SNMO undergoes two distinct magnetic transitions nomenclatured as, (i) T$ _{C} $, the onset in M(T) and (ii) T$ _{d} $, the downturn in M(T). T$ _{C} $, evaluated from the inflection point at FCW M(T) curve (Inset of Fig. \ref{mtmh}(a)), is found to be T$ _{C} $=159.6K. According to GK \cite{GoodenoughKanamoril195559} rule, the virtual hopping of electron from half filled Ni$ ^{2+} $ e$ _{g} $ orbital to empty Mn$ ^{4+} $ e$ _{g} $ orbital in 180$ ^{o} $ geometry should be ferromagnetic in nature. On the other hand, if magnetic exchange is also possible in between Ni e$ _{g} $ and Mn t$ _{2g} $ orbitals, it will be antiferromagnetic. However, in previous theoretical studies on prototype LNMO (for A=La) system, it is observed that t$ _{2g} $ electrons are more localized than e$ _{g} $, the magnetic exchange is governed by Ni-Mn e$ _{g} $ orbital overlap through intermediate O $ p $ orbital and ferromagnetic ground state is energetically more favorable than the antiferromagnetic case \cite{HDas2008, PSanyal2017}. Neutron powder diffraction (NPD) measurements  below T$ _{C} $ on SNMO system reveal long range colinear ferromagnetic ordering of Ni-Mn sublattice moments in $ F_{x}F_{z} $ configurations at cation ordered structures \cite{SMajumder2022prbb}. Therefore, observed magnetic transition at T=T$ _{C} $ is attributed to Ni-O-Mn superexchange interaction driven transformation of paramagnetic to ferromagnetic phase in SNMO. The inverse susceptibility 1/$ \chi $(T) experimental curve starts deviating from Curie-Weiss fitting behaviour at $ \sim $100K above T$ _{C} $ (Inset of Fig. \ref{mtmh}(a)) value, indicating presence of short scale interactions even in paramagnetic state. This is because of cation disorder (Ni-O-Ni) related short range magnetic interactions in the system (discussed later). In past studies conflicting explanations are found regarding the low temperature transition at T=T$ _{d} $. For example, to discuss the origin of downturn behavior it is proposed that there may be, (i) strong magnetocrystalline anisotropy due to spin-orbit coupling between rare earth - transition metal network \cite{RJBooth2009} or (ii) rare earth long range magnetic ordering \cite{WZYang2012} or (iii) reentrant transition to spin glass phase \cite{PNLekshmi2013}. The thermal evolution of microscopic magnetic structure divulges the internal field polarization of paramagnetic Sm moments opposite to ordered Ni-Mn network in the vicinity of T=T$ _{d} $ \cite{SMajumder2022prbb}. Unlike T$ _{C} $, the transition temperature T$ _{d} $ varies with different measuring field strengths due to the competition between internal field and external field acting on the sublattice moments \cite{SMajumder2022prbb, SMajumder2022prbt}. T$ _{d} $, estimated from the temperature point at which low field ($ \mu_{0} $H=25Oe) FCW M(T) curve (data not shown here) achieves maximum moment value, is found to be T$ _{d} $=34.1K. The isothermal magnetization measured as a function of applied magnetic field M(H) at T=10K is shown in Insets (i, ii) of Fig. \ref{mtmh}(b). M(H) curve does not attain saturation even at $ \mu_{0} $H=70kOe of applied field. This is possibly because of opposite arrangement between Sm paramgnetic momments and Ni-Mn ordered ferromagnetic moments or the presence of Ni/Mn cation disorder (discussed later) mediated short scale antiferromagnetic interactions in predominant ferromagnetic ordered host matrix. Thermal evolution of coercive field H$ _{C} $ and moment M$ _{\rm 70kOe} $ are illustrated in Fig. \ref{mtmh}(b). Both of H$ _{C} $ and M$ _{\rm 70kOe} $ behaviors depict abrupt change in the vicinity of T$ _{C} $ and T$ _{d} $. Aforementioned magnetic transition temperatures as obtained from bulk magnetometric experiments have well correspondence with our previous microscopic magnetic structure analysis on SNMO system \cite{SMajumder2022prbb}.

\begin{figure}[t]
\centering
\includegraphics[angle=0,width=1.0\textwidth]{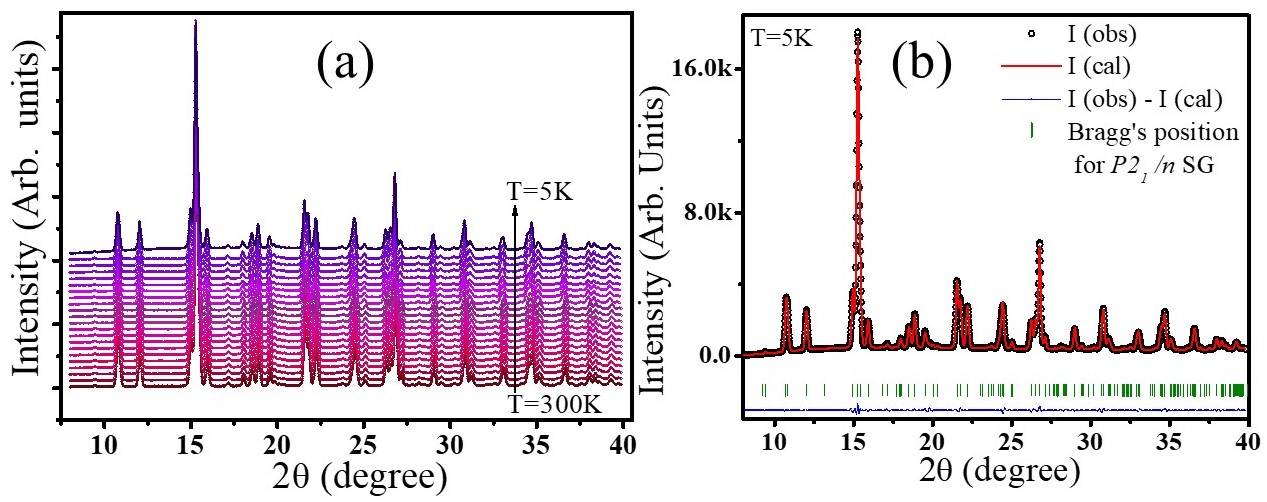}
\caption{(a): Temperature dependency of powder X-ray $ 2\theta $ diffractograms. (b): Representative Rietveld fitting of PXRD $ 2\theta $ scan measured at T=5K showing observed (red open circles), calculated (black solid line) and difference (blue solid line) patterns along with Bragg's positions (green vertical bars).}\label{pxrd}
\end{figure}

\begin{figure}[t]
\centering
\includegraphics[angle=0,width=1.0\textwidth]{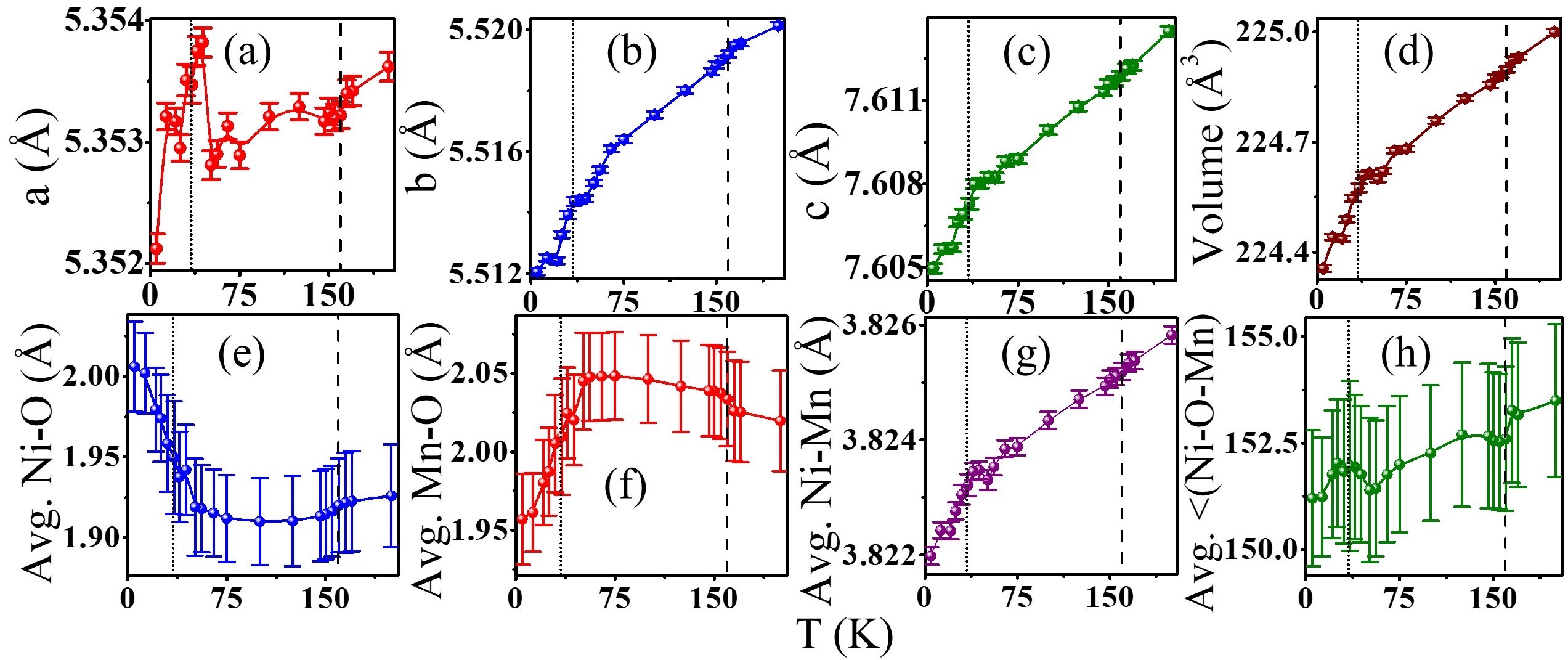}
\caption{The thermal evolution of unit cell parameters (a): a, (b): b, (c): c, (d): volume, atomic bond distances (e): Ni-O, (f): Mn-O, (g): Ni-Mn and (d): average $\langle\overbrace{Ni-O-Mn}$ bond angle obtained from PXRD analysis.}\label{cpbdavst}
\end{figure}

In order to investigate the global structural evolution across observed magnetic transitions, temperature dependent X-ray diffractograms are measured as displayed in Fig. \ref{pxrd}(a). Starting from T=300K down to T=5K, no new Bragg's reflection is observed in PXRD patterns, confirming that crystal symmetry remains the same within the investigated temperature regime. PXRD pattern at T=5K along with Rietveld simulation for monoclinic \textit{P2$_{1}$/n} structure are depicted in Fig. \ref{pxrd}(b). Goodness of fitting indicators for T=5K data analysis are: R$_{p}$=3.11, R$_{wp}$=3.81, R$_{exp}$=4.94, $\chi^2$=0.593. These statistical factors suggest reasonably good agreement between experimentally observed and simulated patterns. The observance of (011) superstructure reflection in NPD profile confirms dominating Ni, Mn cation ordering at respectively \textit{2c}, \textit{2d} sites of SNMO monoclinic \textit{P2$_{1}$/n} (SG 14) lattice \cite{SMajumder2022prbb}. Point to be noted here that it is difficult to resolve such superstructure peak by PXRD measurements owing to nearly equal ionic radius of Ni and Mn species in SNMO. The temperature dependency of a, b, c lattice parameters and corresponding change in unit cell volume are illustrated in Figs. \ref{cpbdavst} (a-d), respectively. At high temperature typical thermal expansion of cell is observed. With lowering temperature, distinguishable abrupt changes from monotonic decreasing trends are found specifically across the magnetic phase transition temperature T=T$ _{d} $, indicating the presence of iso-structural transitions coupled with magnetic ordering in SNMO. As the magnetic exchange interactions depend on the orbital overlap in transition metal - ligand linkage, we have explored the variation of Ni/Mn-O, Ni-Mn average bond lengths and $\langle\overbrace{Ni-O-Mn}$ average bond angle as a function of temperature, which are shown in Figs. \ref{cpbdavst}(a-d), respectively. Bond lengths and bond angel exhibit anomalous behavior in the vicinity of magnetic transition at T$ _{d} $. The co-occurrence of structural and magnetic transitions indicate coupling between crystal structure and magnetic structure in SNMO.

\begin{figure}[t]
\centering
\includegraphics[angle=0,width=1.0\textwidth]{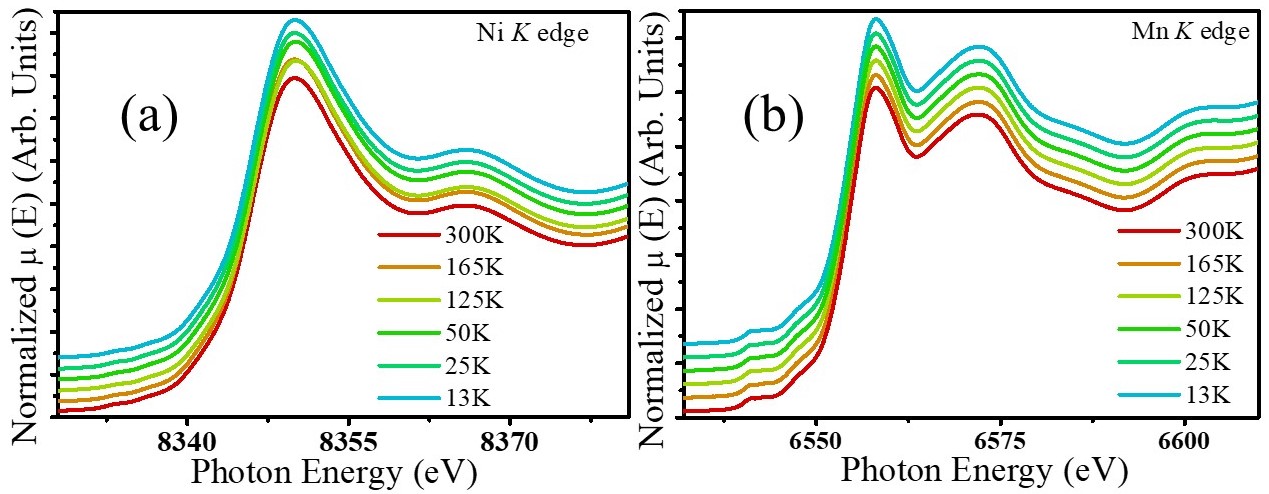}
\caption{Temperature dependent X-ray absorption spectra in near edge region measured across (a): Ni and (b): Mn \textit{K} edges.}\label{xanesvst}
\end{figure}

To check if there is any temperature driven valence state transition of constituent elements liable for the magnetic anomalies in SNMO, temperature dependent X-ray absorption measurements at Ni and Mn \textit{K} edges are carried out. Both Ni and Mn \textit{K} edge XANES show characteristic pre-edge and white line features, as displayed in Figs. \ref{xanesvst}(a, b). The \textit{K} absorption edge is identified as metal 1\textit{s} to empty \textit{p} band electronic transition \cite{FBridges2001}. In general, with increasing valency of absorbing specie, the band edge position of XANES spectra shifts towards higher energy side  \cite{AHdeVries2003}. Here edge energy is estimated from the first inflection point in XANES spectra. Comparing the Ni/Mn \textit{K} XANES spectra for SNMO sample and Mn$ ^{4+} $ (MnO$ _{2} $), Mn$ ^{3+} $ (Mn$ _{2} $O$ _{3} $), Ni$ ^{2+} $ (NiO) and Ni$ ^{3+} $(Ni$ _{2} $O$ _{3} $) standard references (data not shown here) at room temperature, it is confirmed that Ni and Mn have chemical valency in between 2+, 3+ and 3+, 4+, respectively \cite{SMajumder2022prbb}. Noteworthily, both Ni and Mn absorption band edges do not show any apparent shift with respect to measuring temperature variation (Figs. \ref{xanesvst}(a, b)), suggesting that these mixed valence nature remains unaltered across the magnetic transitions.

\begin{figure}[t]
\centering
\includegraphics[angle=0,width=1.0\textwidth]{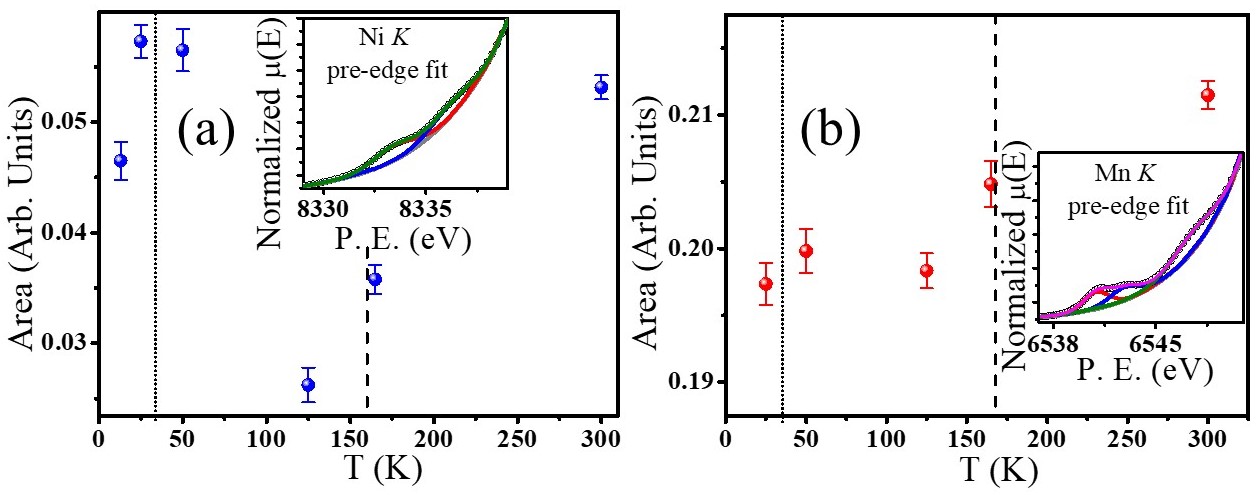}
\caption{Temperature dependency of pre-edge feature integrated intensities for (a): Ni and (b): Mn \textit{K} X-ray absorption spectra. Insets show representative fittings of pre-edge structures recorded at T=300K for Ni and Mn \textit{K} absorption, respectively.}\label{kpreedgevst}
\end{figure}

The pre-edge structures in transition metal $ K $ XANES is possibly originated due to electric quadrupole or / and dipole transitions from metal 1\textit{s} to empty 3\textit{d} states \cite{FBridges2001}. In centrosymmetric structure, $ 1s \rightarrow 3d $ dipole excitation is forbidden. Any deviation from centrosymmetry may introduce metal 3\textit{d}-4\textit{p} hybridization through ligand involvement and then dipole transition will be weakly allowed \cite{FBridges2001}. However, the irreducible representation analysis using group theory calculation predicts that in octahedral system metal 3\textit{d}-4\textit{p} orbital mixing is not allowed \cite{TYamamoto2008}. Point to be noted here that SNMO has centrosymmetric structure (SG: \textit{P2$ _{1} $/n}) where Ni/Mn absorbents are ocahedrally surrounded by O ligands. Therefore, pre-edge structures observed here, are assigned to electric quadrupole transitions from Ni/Mn 1\textit{s} to 3\textit{d} states. It is important to mention that owing to significant overlapping between transition metal 3\textit{d} - O 2\textit{p} orbitals, XANES \textit{K} pre-edge also contains information about the metal - ligand hybridization \cite{FBridges2001}. In a previous report \cite{CGuglieri2011}, it was observed that with varying Mn-O separations in MnO$ _{4} $ tetrahedron, the pre-edge structure of Mn \textit{K} edge absorption changes, which can be understood by considering modification in metal-ligand hybridization due to change in coordination distances. The area under the curve of Ni, Mn pre-edge features are estimated using arc-tangent background function and Gaussian peak shapes as displayed by Insets of Figs. \ref{kpreedgevst}(a, b). The thermal evolution of obtained integrated intensity of Ni, Mn pre-edge curves show anomalous behavior close to T=T$ _{C} $ and T$ _{d} $, as illustrated in Figs. \ref{kpreedgevst}(a, b). These results reveal temperature driven changes in Ni/Mn-O hybridization.

\begin{figure}[t]
\centering
\includegraphics[angle=0,width=1.0\textwidth]{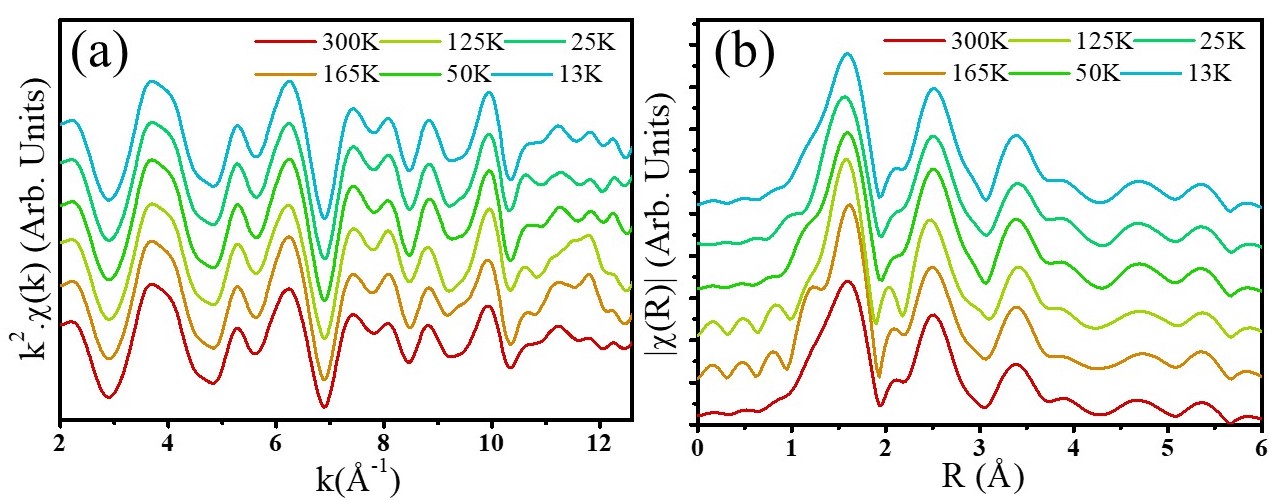}
\caption{The temperature dependencies of extended X-ray absorption (a): $ k^{2} $-weighted $k$-space signals and (b): corresponding modulus of Fourier transformed $R$-space oscillations.}\label{exafsvst}
\end{figure}

To explore the local structural particularities such as coordination environment, short range disorder etc. present in the crystal system, temperature dependent extended X-ray absorption measurements are carried out at Ni \textit{K} edge. Figures \ref{exafsvst}(a, b) present temperature dependent $ k^{2} $-weighted $k$-space oscillations ($ k^{2}\chi(k) $) and corresponding modulus of Fourier transformed $R$-space spectra ($ | \chi (R) |$) respectively, for SNMO Ni \textit{K} edge EXAFS. To probe the Ni/Mn cation disorder present in the system, we have performed quantitative fitting analysis on $ | \chi (R) |$. The theoretical model is generated using cell parameter values as obtained from Rietveld refinement of PXRD data \cite{SMajumder2022prbb}. The fitting range is confined to 1\AA \ $ \leqslant $ R $ \leqslant $ 4.2\AA \ in R-space and 2.5\AA$ ^{-1} $ \ $ \leqslant k \leqslant $ 12\AA$ ^{-1} $ in $k$-space. Within this selected region, EXAFS signal is contributed from photoelectron scattering by O anions at nearest neighbor octahedral sites, Sm cations at second nearest neighbor sites and Mn/Ni cations at next B-sites connected with core absorber Ni ion through intermediate O ion. The most appropriate single scattering and multiple scattering paths are employed in fitting model to adequately generate the experimental spectral behavior. Although single scattering in Ni-O, Ni-Sm and Ni-Mn/Ni linkages have dominating contribution, the multiple scattering (forward triangle geometry) effects through Ni-O-Mn/Ni can not be ignored in analysis.

\begin{figure}[t]
\centering
\includegraphics[angle=0,width=1.0\textwidth]{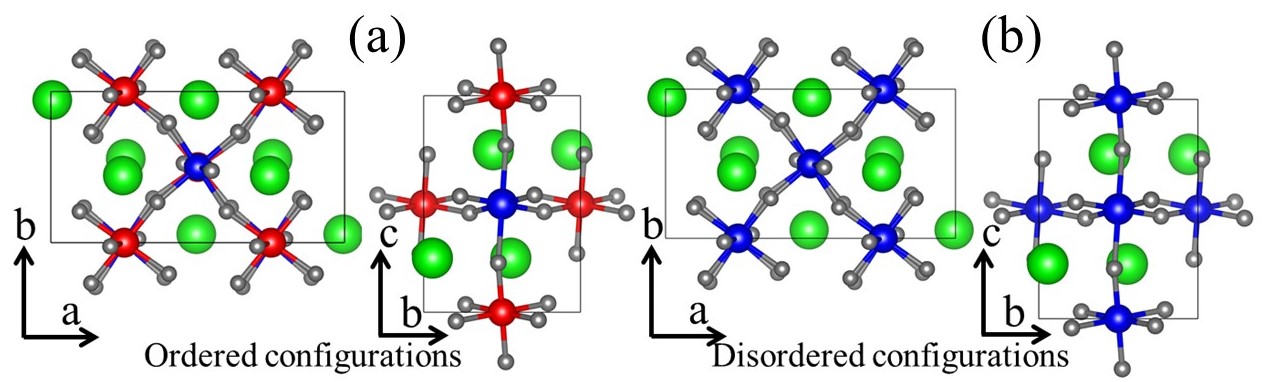}
\caption{Schematic representations of cation (a): ordered (Ni-O-Mn) and (b): disordered (Ni-O-Ni) unit cells showing a-b, b-c crystallographic planes. Green, Blue, Red and Grey spheres are for Sm, Ni, Mn and O ions, respectively.}\label{orddisstr}
\end{figure}

To account for Ni/Mn anti-site disorder in the fitting model, two different structures, both having same core absorber as Ni atom, are considered, as illustrated in Figs. \ref{orddisstr}(a, b). In ideal ordered configuration, all the next B-sites are filled with Mn atoms (Fig. \ref{orddisstr}(a)). Whereas, in ideal disordered case, all the next B-sites are occupied with Ni atoms (Fig. \ref{orddisstr}(b)). The simulated EXAFS spectra have a weighted convolution from these two types of cells \cite{BNRao2016}. Ni-Mn and Ni-Ni shell coordination numbers are refined to evaluate the fractional weights of each structures. The concentration of anti-site disorder is quantified in terms of the probability of encountering disordered bond configurations Q$_{ASD} $, defined as\cite{CMeneghini2009}, 
\begin{eqnarray}\label{EqQASDE}
Q^{XAS}_{ASD} & = & N_{Ni-Ni} / N_{B-site} \nonumber\\
N_{B-site} & = & N_{Ni-Mn} + N_{Ni-Ni}
\end{eqnarray}
where, N$_{Ni-Mn}$, N$_{Ni-Ni}$ are the coordination numbers for ordered and disordered structures respectively and N$_{B-site}$ is the total coordination number of all (Mn/Ni) B-site configurations. 

\begin{figure}[t]
\centering
\includegraphics[angle=0,width=1.0\textwidth]{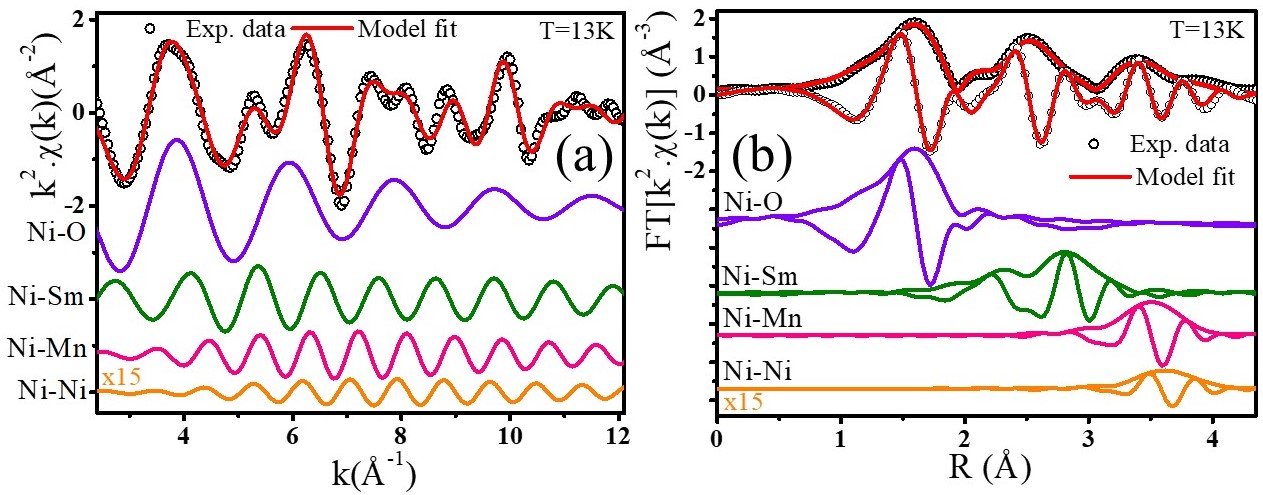}
\caption{Representative fitting of extended X-ray absorption (a): $ k^{2} $-weighted $k$-space signal and (b): corresponding Fourier transformed $R$-space oscillation recorder at T=13K. Spectral contributions of different coordination shells accounted in fitting model, are shifted in vertical axes to have clear visualizations.}\label{exafsfit}
\end{figure}

In fitting model the amplitude reduction factor (s$_{0}^{2}$) is fixed at 0.84. This value is obtained by analyzing Ni \textit{K} EXAFS spectra for NiO. As s$_{0}^{2}$ is chemically transferable, same value can be used for SNMO case also. Equal energy shift ($ \Delta $E$ _{0} $) is used for all coordination shells and for all temperatures to avoid any fictitious artifact due to correlation with coordination distances. During refinement process, the average coordination distances (R) and mean-square relative displacement (MSRD) factors ($ \sigma^{2} $) are freed to refine. In all fitting, the total coordination numbers (N) are kept fixed at corresponding crystallographic values, N$_{O}$=6, N$_{Sm}$=8 and N$_{B-site}$=6. This should eliminate spurious correlation with $ \sigma^{2} $. Some preliminary trial refinements suggest that same $ \sigma^{2} $ can be shared for Ni-Mn/Ni scattering paths as they are highly correlated. The quality of fitting is monitored by $ R $, factor defined as, 

\begin{equation}\label{EqRfactor}
R = \frac{\sum_{i} [Re(\chi_{d}(R_{i})-\chi_{t}(R_{i}))^{2}+Im(\chi_{d}(R_{i})-\chi_{t}(R_{i}))^{2}]}{\sum_{i} [Re(\chi_{d}(R_{i}))^{2}+Im(\chi_{d}(R_{i}))^{2}]}
\end{equation}
where, $ \chi_{d} $ and $ \chi_{t} $ correspond to the experimental and theoretical $ \chi(R) $ values, respectively. 

The best fit along with experimentally observed EXAFS pattern measured at T=13K are displayed by $ k^{2} . \chi (k)$ and corresponding Fourier transform of $ k^{2} . \chi (k)$, as illustrated in Figs. \ref{exafsfit}(a, b), respectively. Here, the partial contribution from different coordination shells used in model analysis are vertically shifted to have better visualization. The goodness of fitting indicator $ R $ having value 0.009 suggests reasonable agreement between experimental and simulated profiles. The estimated anti-side disorder concentration Q$_{ASD}$ is found to be 3.6(3)$ \% $. Obtained Q$_{ASD}$ value is in well agreement with NPD analysis \cite{SMajumder2022prbb}. Therefore, EXAFS modeling confirms the presence of Ni/Mn cation ordered and disordered structures in SNMO lattice. Such local anti-site disorder lefts its imprint in magnetization behaviors by introducing short scale antiferromagnetic interaction in background of long range ferromagnetic ordered SNMO lattice \cite{SMajumder2022prbb, SMajumder2022prbt, SMajumder2022jpcm}. 

\begin{figure}[t]
\centering
\includegraphics[angle=0,width=1.0\textwidth]{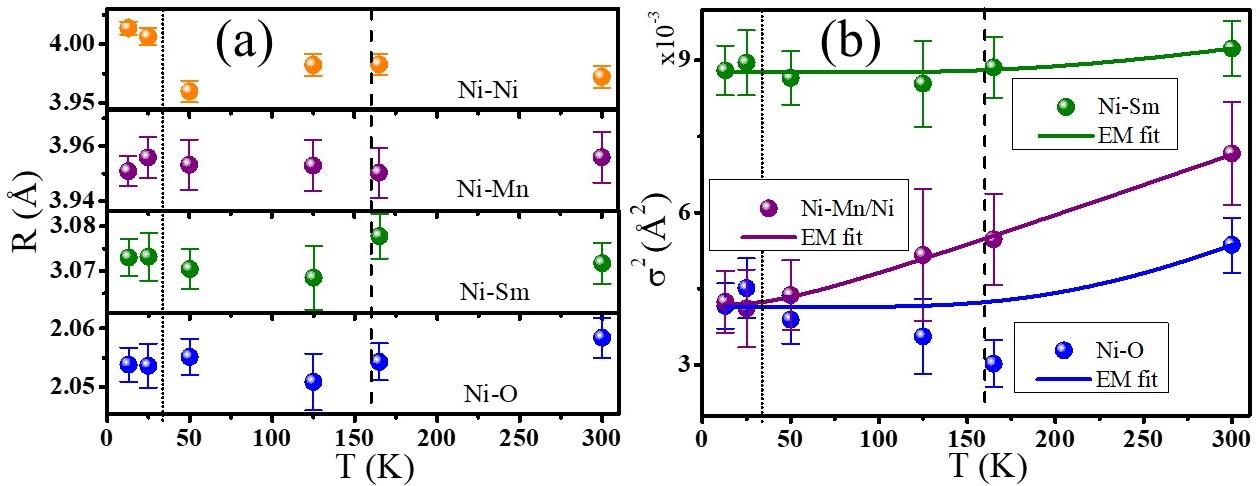}
\caption{Thermal evolution of (a): coordination distances and (b): corresponding bond disorder parameters obtained from EXAFS analysis. In (b) geometric symbols are for MSRD values estimated from EXAFS studies and solid lines are for the Einstein model behaviors.}\label{exafsp}
\end{figure}

The thermal evolution of R and $ \sigma^{2} $ obtained from EXAFS fitting at different temperature values, are displayed in Figs. \ref{exafsp}(a, b), respectively. For all coordination shells, $ \sigma^{2} $ show increasing trend with rising measurement temperatures, indicates more thermal agitation at elevated temperatures. The temperature dependencies of $ \sigma^{2} $ can be described considering correlated Einstein model where vibrations in atomic bonds are assumed as harmonic oscillators with single frequency $ \omega_{E} $, known as the Einstein frequency \cite{SMahana2018}. According to Einstein model $ \sigma^{2}(T) $ can be expressed as \cite{SMahana2018, ASurampalli2019}, 
\begin{eqnarray}\label{EqQASDE}
\sigma^{2}(T) & = & \sigma_{s}^{2} + \frac{\hbar}{2\mu\omega_{E}} {\rm coth}\left(\frac{\hbar\omega_{E}}{2k_{B}T}\right), \nonumber\\
\frac{\hbar\omega_{E}}{k_{B}} & = & \theta_{E}
\end{eqnarray}
where, $ \sigma_{s}^{2} $ and $ \mu $ represent static bond disorder contribution and reduced mass of atomic bond, respectively. $ \theta_{E} $, known as the Einstein temperature, measures atomic bond stiffness. Obtained $ \theta_{E} $ values reduce as $ \theta_{E}^{Ni-O}(\sim$872K$) > \theta_{E}^{Ni-Sm}(\sim$779K$) > \theta_{E}^{Ni-Mn/Ni}(\sim$124K$)$, suggesting decrease in phonon activation energies with increasing coordination distances \cite{ASurampalli2019}. Among all possible bonding linked with Ni core, the largest phonon activation energy for Ni-O atom pairs indicates highest rigidity of NiO$ _{6} $ octahedra \cite{SMahana2018}. Noteworthily, experimentally obtained $ \sigma^{2}(T) $ for Ni-O bond deviates from the modeled behavior at T=T$ _{C} $ by approximately 29$ \% $ which is larger than statistical uncertainty and hence is not associated to any artifact. Therefore, observed change in $ \sigma^{2}(T)_{Ni-O} $ points out the temperature driven anomalous variation of bond disorder across the magnetic transition.

\begin{figure*}[t]
\centering
\includegraphics[angle=0,width=0.45\textwidth]{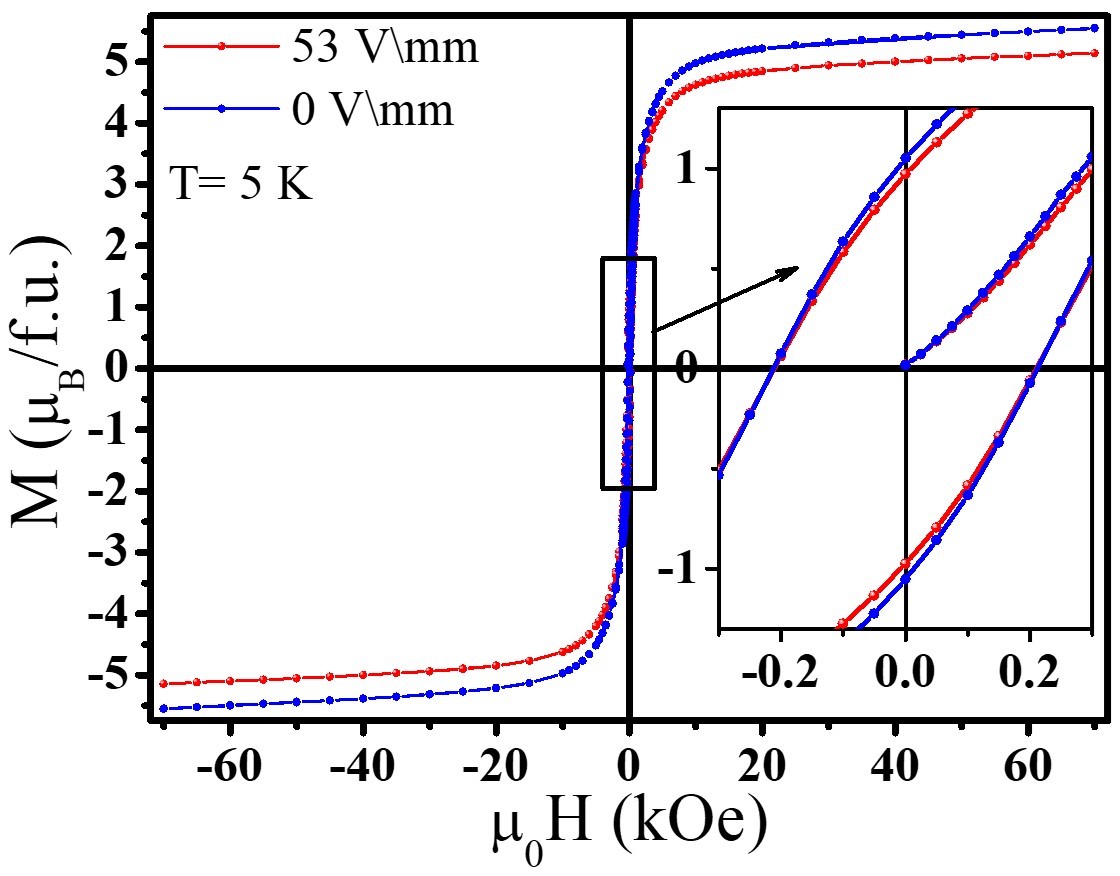}
\caption{\textit{Magnetic isotherms at T = 5 K in absence and presence of applied electric field bias of 53 V/mm. Inset: Enlarged view across the low field region of the M(H) curve.}}\label{msvsefield}
\end{figure*}

The local and global crystal structure analysis utilizing temperature dependent EXAFS and PXRD measurements reveal presence of isostructural changes in SNMO. The variation in crystal environment drives modification in metal - ligand hybridization which is evidenced from electronic structure studies using thermal evolution of XANES results. Alteration in lattice structure and electronic energy landscape is liable for the observed magnetic transitions in SNMO. In order to probe the electro-magnetic correlation in the SNMO system, we have measured the isothermal magnetization in the presence and absence of applied electric filed, as depicted in Fig. \ref{msvsefield}. The strength of electro-magnetic coupling, estimated as $ \frac{(M(0)-M(E))}{M(0)} \times 100 $, is found to be 7.3$ \% $ at T = 5 K and E = 53 V/mm. Therefore, our findings present conclusive evidences of the coupled structural, electronic and magnetic phases in SNMO double perovskite. Such couplings are desirable from device application view point owing to have additional degrees of freedom to control functional aspects of the system.

\section{CONCLUSION}
In summary, we have explored temperature dependency on structural, electronic and magnetic properties of Sm$ _{2} $NiMnO$ _{6} $ double perovskite. From bulk magnetization measurements it is observed that SNMO encounters two magnetic transitions, attributed to ferromagnetic ordering of Ni-Mn network and polarazation of Sm paramagnetic moments opposite to Ni-Mn sublattice, at T$ _{C} $=159.6K and T$ _{d} $=34.1K, respectively. SNMO crystallizes in \textit{P2$ _{1} $/n} monoclinic structure with dominating Ni/Mn cation ordered phases. X-ray diffraction results exhibit isostructural change in global crystal structure in the vicinity of T$ _{C} $ and T$ _{d} $, whereas the lattice symmetry remains the same within investigated temperature regime (5K$ \leqslant $T$ \leqslant $300K). Moreover, the average bond distances and bond angle between Ni, Mn cations via intermediate O anion are observed to show abrupt variation across magnetic transition, particularly near T=T$ _{d} $. The local coordination environment studies using Ni \textit{K} edge extended X-ray absorption spectroscopy reveal deviation of Ni-O bonding feature from expected Einstein model behavior particularly at T$ _{C} $. Furthermore, existence of Ni/Mn anti-site disorder is confirmed which influences the bulk magnetic characters. Absence of temperature dependent valence state transition of Ni, Mn ions is confirmed from X-ray absorption near edge spectroscopy. However, changes in Ni, Mn 3\textit{d} - O 2\textit{p} hybridization around T$ _{C} $ and T$ _{d} $ are evidenced. The temperature driven alteration in local as well as global structures provokes modification in energy landscape which eventually results in variation of magnetic observable of SNMO system. Combining these results, we have demonstrated a ferromagnetic insulator platform where structural, electronic and magnetic properties are coupled with each other. We hope present work will have huge relevance on tuning functional aspects of all prototype double perovsite system by manipulating any of the aforementioned degrees of freedom. 
\\
\section*{ACKNOWLEDGMENTS}
Authors gratefully acknowledge Dr. Anil Kumar Sinha (RRCAT, India) for his help in PXRD measurements. Thanks to Indus Synchrotron RRCAT, India and Elettra-Sincrotrone, Italy for giving access to experimental facilities. Authors acknowledge the Department of Science and Technology, Government of India; Indian Institute of Science, Italian Government and Elettra for providing financial support through Indo-Italian Program of Cooperation (No. INT/ITALY/P-22/2016 (SP)) to perform experiments at Elettra-Sincrotrone. S.M. thanks Mr. Akash Surampalli (UGC DAE CSR, India) for fruitful discussions.

\section*{REFERENCES}
\bibliography{}

\end{document}